\documentstyle[12pt,epsfig]{article}
\newcommand{\bear}{\begin{eqnarray*}}
\newcommand{\eear}{\end{eqnarray*}}
\newcommand{\be}{\begin{equation}}
\newcommand{\ee}{\end{equation}}

\newcommand{\cD}{{\cal D}}
\newcommand{\cE}{{\cal E}}
\newcommand{\wt}{\mbox{wt}}
\newcommand{\Supp}{\mbox{Supp}}
\newcommand{\proof}{{\bf Proof }}
\newcommand{\qed}{\hfill$\Box$}
\newcommand{\Tr}{\mbox{Tr}}
\newcommand{\Id}{\mbox{Id}}
\newcommand{\Res}{\mbox{Res}}

\begin{document}
\title{Asymptotically Good Quantum Codes}

\author{
Alexei Ashikhmin
\thanks{Bell Laboratories, Lucent
Technologies, 600 Mountain Avenue, Murray Hill, NJ 07974, USA,
{\tt aea@research.bell-labs.com}}
\and
Simon Litsyn \thanks{Department of Electrical
Engineering--Systems, Tel Aviv University, Ramat Aviv
69978, Israel, {\tt litsyn@eng.tau.ac.il}}
\and Michael A. Tsfasman
\thanks{Institute for Information Transmission Problems,
 Russian Academy of Sciences,
 19 Bolshoi Karetny, 101447 Moscow GSP-4,
 Russia, {\tt tsfasman@iitp.ru}}
}
\date{}
\maketitle
\newcommand{\Bbb}{\bf}
\newtheorem{theorem}{Theorem}
\newtheorem{corollary}{Corollary}
\newtheorem{lemma}{Lemma}

\begin{abstract}
\noindent
{U}sing algebraic geometry codes we give a polynomial construction of
quantum codes with asymptotically non-zero rate and relative distance.
\end{abstract}

\section{Introduction}
Let ${\cal B}={{\bf C}^2}$, an element of ${\cal B}$ is called a {\it qubit}.
The space ${\cal B}^n={\cal B}^{\otimes n}=({\bf C}^2)^{\otimes n}$ 
is the space of quantum words of length $n$. 
An $((n,K))$ {\it quantum code} $Q$ is a $K$-dimensional linear subspace of 
${\cal B}^n$.
The parameters $n$ and $K$ are called the {\it length} and
the {\it size} (or {\it cardinality}) of the  code. 

Let ${\bf L}({\cal B}^n)$ be the space of linear operators on ${\cal B}^n$.
A {\it quantum information message} is
a vector $w \in Q$. The message 
$w$ can be altered  by a linear operator $E\in {\bf L}({\cal B}^n)$, called
an {\it error operator}. 

Let us define the set $\Supp E \subseteq [1,n]$ in the following way. 
Consider the action of $E$ on ${\cal B}^n$. If $E$
can be written as $\Id_j \otimes E'$, 
where $\Id_j$ is the identity operator acting on the $j$-th tensor 
component and 
$E'$ an operator on the tensor product of the other components, then 
$j\notin \Supp E$. 
The {\it weight} of $E$ is defined as $\wt(E)=\left|\Supp E\right|$.

We say that $E$ is {\it detectable} by $Q$ 
if for any two $v,u\in Q$ if $v\perp u$ 
then $v\perp E(u)$.
Let $d_Q$ be the maximum integer such that $Q$ can detect any error of 
weight $d_Q-1$ or less; 
$d_Q$ is called the {\it minimum distance} of $Q$. We say 
that $Q$ is an
$((n,K,d_Q))$-code. It can be proved that the code $Q$ can correct
any error of weight $\lfloor {d_Q-1\over 2}\rfloor$ or less.

\noindent {\bf Remark}  One can find  a
more details discussion of the notions of quantum minimum distance,
quantum detection, and quantum correction
in  \cite{ashi99a},  \cite{ashi99c}, \cite{ksv99}, \cite{knil97}.

Probably the most interesting and important class of
quantum codes are quantum stabilizer codes. 
These codes
can be viewed as natural analogues of classical linear codes. 
To define a quantum stabilizer code we first introduce another
class of (non-quantum) codes.

Let $T={\bf F}_4$. The non-trivial automorphism of ${\bf F}_4$ over  
${\bf F}_2$
is called complex conjugation and denoted in the same way. We fix a 
(symplectic) form on $T^n$ given by $\omega (x,y)=\Tr(x{\bar y})$. 
There is a usual ${\bf F}_4$ Hamming norm on $T^n$.
A {\it small symplectic code} $F\subset T^n$ is an $\omega$-isotropic 
${\bf F}_2$-subspace in $T^n$, i.e., $\omega(x,y)=0$ for any $x,y\in F$. 
Its {\it minimal distance} $d=d_F$ is defined as the minimum 
${\bf F}_4$ Hamming norm of a non-trivial vector in $F$. 
Its dimension $k=k_F$
is its ${\bf F}_2$-dimension, in particular, $k\le n$. The $\omega$-dual 
$F^{\omega}$ of a small
symplectic code $F$ is called a {\it large symplectic code}, 
for a large symplectic code 
we have $n\le k_{F^{\omega}} \le 2n$. Of course, $F\subset F^{\omega}$.

Let $F\subset T^n$ be a small symplectic code with parameters $[n,k,d]$. 
We are going to define the
standard stabilizer code $Q_F\subset {\cal B}^n$ corresponding to $F$. 
Let ${\bf F}_4=\{0,1,\varepsilon,\bar\varepsilon\}$. Set
$$
\sigma(0)=\left[ \begin{array}{cc}
1 & 0 \\
0 & 1
\end{array} \right],
\sigma(\varepsilon)=\left[ \begin{array}{cc}
0 & 1 \\
1 & 0
\end{array} \right],
\sigma(\bar\varepsilon)=\left[ \begin{array}{cc}
1 & 0 \\
0 & -1
\end{array} \right],
\sigma(1)=\left[ \begin{array}{cc}
0 & -i \\
i & 0
\end{array} \right].
$$
These are the usual {\it Pauli matrices}. 
Then, for $t=(t_1,\ldots,t_n)\in T^n$
we put 
\begin{equation}\label{eqn:map}
\sigma(t)=\sigma(t_1)\otimes\dots\otimes\sigma(t_n).
\end{equation}
 We get a
map (of sets) $\sigma:T^n\rightarrow {\bf L}({\cal B}^n)$.
Being restricted to a small symplectic code $F\in T^n$, the map $\sigma$
happens to be almost a group homomorphism, namely for $f_1,f_2\in F$ we have
$$\sigma(f_1)\sigma(f_2)=\sigma(f_2)\sigma(f_1)=\pm\sigma(f_1+f_2),$$
in particular $\sigma(f_1)$ and $\sigma(f_2)$ commute. This makes it possible
to consider the subspace of ${\cal B}^n$ fixed by $\sigma(F)$ in the 
following way.
Let ${\cal F}= \{f_1,\dots,f_k\}$ be an ${\bf F}_2$-basis of $F$ and let
$\mu=\{\mu_1,\dots,\mu_k\}$, $\mu_i\in\{\pm 1\}$.

Define
$Q_{{\cal F},\mu}$ as follows
$$
Q_{{\cal F},\mu}=\{x\in{\cal B}^n\;\mid\; \sigma(f_i)(x)=\mu_i x 
\mbox{ for any } i=1,\dots,k\}.
$$

The quantum code $Q_{{\cal F},\mu}$ is called a {\it stabilizer code}.
For any $f\in F$ the operator $\sigma(f)$ acts on $Q_{{\cal F},\mu}$
as $\pm 1$. 

The small symplectic code $F$ being fixed, we get $2^k$ different codes
$Q_{{\cal F},\mu}$. Their properties, we are interested in, do not depend
on the choice of $\cal F$ and $\mu$, and by abuse of notation we call 
each of them $Q_F$.

The main theorem on stabilizer codes says that the parameters of the 
obtained quantum codes are
\begin{equation}
\label{qp}
K_{Q_{F}}=2^{n-k_F}\;\;\;,
d_{Q_{F}}=\min_{f\in {F^{\omega}}\setminus F}\| f\|\ge d_{F^{\omega}}\;.
\end{equation}

\noindent {\bf Remark}  Detailed descriptions of quantum
stabilizer codes including the proof of the above statements on their
parameters can be found in 
\cite{cald96}, \cite{gott97}, \cite{ksv99},
\cite{shor95}, \cite{stea96a}.

Let $k_Q=\log_2 K_Q$ and set
$$
R_Q= {k_Q \over n} \mbox{ and } \delta_Q={ d_Q \over n}.
$$
We are interested in
$$
R(\delta)= \limsup_{n \rightarrow \infty} R_Q,
$$
where the limit is taken over all codes with $\delta_Q
\ge \delta$.

The best known {\it nonconstructive} lower bound on $R(\delta) $ was obtained in \cite{cald98}
via codes over ${\bf F}_4$:
\begin{equation}
\label{vg4}
R(\delta) \ge 1-\delta \log_2 3-H(\delta) ,
\end{equation}
where
$H(x)= -x \log_2 x - (1-x) \log_2 (1-x)$ is the binary entropy function.
For upper bounds see \cite{ashi99c}.

Several methods  were proposed to construct quantum codes, see, e.g.
\cite{cald97}, \cite{cald98}, \cite{cald96}, \cite{cohe00},
\cite{gott96}, \cite{knil97}, \cite{shor95}, \cite{stea96a}, 
\cite{stea96b}, \cite{stea98}.
However, when $n$ grows for a fixed $R>0$,
the relative minimum distance $\delta$ of all these codes tends to zero.

In this paper we give a (polynomial in $n$) construction of quantum 
codes from
algebraic geometry codes, so that in a certain interval of rates 
$R$ the relative minimum distance of these quantum codes is
separated from zero, i.e., we construct a family of {\it asymptotically 
good quantum codes}.

The construction proceeds in four steps. Algebraic curves give us
asymptotically
good nonbinary algebraic geometry codes, and we provide that each of 
them contains its dual. 
Then we take a binary symbolwise expansion in a self-dual basis
of the codewords of these algebraic geometry codes, 
so that the resulting binary codes
also contain their duals. Then we plug these codes into Steane's construction 
\cite{stea98} to construct good symplectic codes. The corresponding 
quantum codes are  asymptotically good.

To make the exposition simpler, we follow this path backwards. 
We have already explained how quantum codes are related to symplectic codes.
In Section \ref{sec:BO} we recall Steane's construction of symplectic codes 
starting from
triples $D'\supset D \supset D^\perp$ of binary codes. Section \ref{sec:NB}
explains
how to construct binary codes containing there duals from 
codes over ${\bf F}_{2^m}$ with the same property. In Section \ref{sec:AG} 
we produce
necessary algebraic geometry codes. Finally, in Section \ref{sec:Q} 
we sum up to get
the parameters. Here is the result (see Fig.1).

\begin{theorem}
\label{main}
For any $\delta\in (0,\frac{1}{18})$ and $R$ lying on the broken line given
by the piecewise linear function 
\[
R(\delta)=1-{2\over 2^m-2}-{10\over 3}m\delta \;\mbox{ \rm for }
\delta\in [\delta_{m},\delta_{m-1}]\;,
\]
where $m=3,4,5,\ldots$; $\;\delta_2=\frac{1}{18}$ and 
\[
\delta_m={3\over 5}{2^m \over (2^m-2)(2^{m+1}-2)}\; \mbox{ \rm for }
m=3,4,5,\ldots,
\]
there exist polynomially constructible families of quantum codes with 
$n\rightarrow \infty$ and asymptotic parameters greater than or equal to 
$(\delta,R)$.
\end{theorem}

\section{From binary codes to symplectic codes}
\label{sec:BO}
We follow Steane's construction \cite{stea98} 
with improved estimates on the parameters 
given by
by Cohen, Encheva and Litsyn \cite{cohe00}.

We start with a triple $D'\supset D \supset D^\perp$ of binary codes,
where $D$ is an $[n,k,d]$-code containing its dual $D^\perp$, 
and $D'$ a larger $[n,k']$-code with $k'\ge k+2$. 
Let $G$ be
a generator matrix of $D$, and let $G'$ be such a matrix that
$$
\left( \begin{array}{c}  G \\ G' \end{array} \right)
$$
is a generator matrix of $D'$.
Denote by $d'_2$ the second generalized weight of $D'$, 
i.e., the minimum weight of the bitwise OR
of two different nonzero codewords 
(see \cite{ashi99b}, \cite{cohe94}, \cite{tsfa95}, \cite{wei91}
for properties and known bounds).
Form the code
$C\subset {\bf F}_2^{2n}$ with the generator matrix
$$
\left( \begin{array}{cc} G & 0 \\
                           0 & G \\
                           G'& G''
\end{array}
\right),
$$
where the matrix $G''$ is obtained from $G'$ by permuting its rows so that
no row stays on its place.

Fix the following
${\bf F}_2$-linear isomorphism between ${\bf F}_2^{2n}$ and 
${\bf F}_4^n$ first mapping 
$(x_1,\ldots,x_n, y_1,\ldots,y_n)\in {\bf F}_2^{2n}$ to
$((x_1,y_1),\ldots,(x_n,y_n))\in ({\bf F}_2^2)^n$ and then identifying
${\bf F}_2^2$ and ${\bf F}_4$ by $(0,0)=0, (0,1)=\varepsilon, 
(1,0)=\bar\varepsilon, (1,1)=1$.
The image of $C$ under this map is $F\subset {\bf F}_4^n$.
Here is an estimate for its parameters \cite{stea98}, \cite{cohe00}: 

\begin{theorem} \label{th:st1} 
The code $F\subset{\bf F}_4^n$ is a large symplectic code, i.e.,
$F\supset F^\omega$. Its parameters are $k_F=k+k'$ and 
$d_F\ge \min \left(d, d'_2\right)$. 
\end{theorem}

\proof  
Let $x=(a_1,\ldots,a_n,b_1,\ldots,b_n)$ and 
$x'=(a'_1,\ldots,a'_n,b'_1,\ldots,b'_n)$.
We choose the above
identification between ${\bf F}_4^n$ and ${\bf F}_2^{2n}$. 
In the basis of ${\bf F}_2^{2n}$ the form $\omega(x,x')$ is given by
$\omega(x,x')=\sum_{j=1,\ldots,n}a_j b'_j+a'_j b_j$.
Then suppose that $x\in F^\omega$. This means that $\omega(x,x')=0$
for any $x'\in F$. In particular, this is true for 
$x'=(a'_1,\ldots,a'_n,0,\ldots,0)$ and 
$x'=(0,\ldots,0,b'_1,\ldots,b'_n)$. We get
$\sum_{j=1,\ldots,n}a_j b'_j=0$ for any $(b'_1,\ldots,b'_n)\in D$, 
and therefore 
$(a_1,\ldots,a_n)\in D^\perp\subset D$. Analogously, 
$(b_1,\ldots,b_n)\in D^\perp\subset D$, and we see that $x\in F$.

The value of $k_F$ is obvious. Then we have to estimate $d_F$. Let
$x\in F$. Then 
$$
x=(a_1,\ldots,a_n,0,\ldots,0)+(0,\ldots,0,b_1,\ldots,b_n)
+(a'_1,\ldots,a'_n,b'_1,\ldots,b'_n),
$$
where $(a_1,\ldots,a_n)\in D$, $(b_1,\ldots,b_n)\in D$, and 
$(a'_1,\ldots,a'_n,b'_1,\ldots,b'_n)\in D'$. If the last summand is zero,
the number of non-zero pairs $(a_j,b_j)$ is at least $d$. If it is non-zero,
then both $(a_1+a'_1,\ldots,a_n+a'_n)$ and $(b_1+b'_1,\ldots,b_n+b'_n)$ lie
in $D'$ and they are different since two generators of $D'$ not lying in
$D$ cannot differ by an element of $D$. Hence, the ${\bf F}_4$-weight of
the sum is at least $d'_2$.
\qed

\begin{corollary}\label{c:st1}
The parameters of the corresponding quantum stabilizer code $Q_F$ 
satisfy 
$$k_{Q_F}= k + k' - n,\;\;, d_{Q_F}
\ge \min \left(d, d'_2 \right)\ge  \min \left(d, {3\over 2}d' \right).$$
\end{corollary}

\proof
By (\ref{qp}) the dimension $k_{Q_F}=n-k_{F^\omega}=n-(2n-k_F)=k+k'-n$.
The first inequality is also that of (\ref{qp}). 

To prove that $d'_2 \ge {3\over 2}d'$ write 
two different vectors one below the other.
Let the number of columns
$(0,0)$, $(0,1)$, $(1,0)$, $(1,1)$ equal, respectively, $a_1$, $a_2$,
 $a_3$, $a_4$.
Then $d'_2=a_2+a_3+a_4$. The weight of the first vector is $a_3+a_4\ge d'$,
of the second $a_2+a_4\ge d'$, and of their sum $a_2+a_3\ge d'$. Summing up
we get the result.
\qed

To apply this construction one needs good binary codes 
with $D^{\perp} \subset D$.

\section{From non-binary to binary codes}
\label{sec:NB}
The following theorem is due to T.Kasami an S.Lin \cite{kasami98}.

\begin{theorem} 
\label{th:expan}
Let $C$ be a code over ${\bf F}_{2^m}$ and $C^{\perp} \subset C$.
Let $\alpha_i,i=1,\ldots ,m$ be a self-dual basis of ${\bf F}_{2^m}$ over
${\bf F}_{2}$,
i.e.,
$$
\Tr(\alpha_i\alpha_j)=\delta_{ij}.
$$
Let $D$ and $D^\perp$ be codes
obtained by the symbolwise binary expansion of codes $C$ and $C^\perp$ in
the basis $\alpha_i$. Then
$D^\perp \subset D$ and
$D^\perp$ is the binary dual of $D$.

\end{theorem}

\proof
The first statement is obvious.

Let us prove the second one.
Let ${x}=(x_1,x_2,\ldots ,x_n)\in C$ and
${y}=(y_1,y_2, \ldots ,y_n)\in C^\perp$.
Let
$$
x_j=\sum_{i=1}^{m} x_i^{(j)} \alpha_i\;, 
$$
$$
y_j=\sum_{i=1}^{m} y_i^{(j)} \alpha_i\;.
$$
Then
$$
 \sum_{j=1}^n x_jy_j={\bf x}{\bf y}=0
$$
Hence
\bear
0 & = &\Tr\left(\sum_{j=1}^n x_jy_j\right) \\
& = &\Tr\left(\sum_{j=1}^n \sum_{i=1}^{m}\sum_{t=1}^{m}
x_i^{(j)}y_t^{(j)}\alpha_i
\alpha_t\right)\\
& = &\sum_{j=1}^n \sum_{i=1}^{m}\sum_{t=1}^{m}
x_i^{(j)}y_t^{(j)} \Tr\left(\alpha_i
\alpha_t\right)\\
& = &\sum_{j=1}^n \sum_{i=1}^{m} x_i^{(j)}y_i^{(j)}.
\eear
So we have proved that $(D)^\perp\supseteq D^\perp$. It rests to remark that
the dimensions of $D$ and $D^\perp$ are complimentary.
\qed

Of course, if we start from a triple $C'\supset C \supset C^\perp$ 
of codes over ${\bf F}_{2^m}$ the same descent gives us a
triple $D'\supset D \supset D^\perp$ of binary codes.

\section{From algebraic curves to codes}
\label{sec:AG}

In this section we follow standard algebraic geometry constructions presented 
in \cite{tsvl91}, proving that they satisfy some extra 
properties needed to use them in above constructions. 
Namely, we want a triple $C'\supset C \supset C^\perp$ 
of codes over ${\bf F}_{2^m}$ with good parameters. 
Let us start from looking for algebraic codes containing their duals.

Let $w\in ({\bf F}_q^*)^n$. For a code $C\subset{\bf F}_q^n$ we define
\[
C_w^\perp=\left\{ x\in {\Bbb F}_q^n:\sum w_ix_iy_i=0 \mbox{
for any }y\in
C \right\} .
\]

Let $X$ be a (smooth projective geometrically irreducible algebraic) curve
of genus $g$ defined over ${\Bbb F}_q$, let $\cD$ be an effective divisor of
degree $a$ and ${\cal P'}=\left\{ P_1,\ldots ,P_{n'} \right\} 
\subseteq X\left(
{\Bbb F}_q\right)$ a set of ${\Bbb F}_q$-points such that $\Supp\cD\cap
{\cal P'}=\emptyset$; we set ${\bf P'}=P_1+\ldots +P_{n'}$. As usual,
\[
L\left( \cD\right) =\left\{ f\in {\Bbb F}_q\left( X\right) :\left( f\right)
+\cD\geq 0\right\} \cup \left\{ 0\right\}
\]
is the space of functions associated to the divisor, and
\[
\Omega \left( \cD\right) =\left\{ \omega \in {\Bbb F}_q\left( X\right) 
:\left(
\omega \right) +\cD\geq 0\right\} \cup \left\{ 0\right\}
\]
that of differential forms.

Suppose that $a\le\frac {n'}2+\frac g2-1$, then 
for any effective
divisor $\cE$ of degree $\deg \cE=n'+g-2-2a$ 
we have $\deg\left(K+{\bf P'}-2\cD-\cE\right)=g$ and by the Riemann--Roch
theorem 
there exists an 
$\omega \in\Omega\left({\bf P'}-2\cD-\cE\right)$. Unfortunately, working 
over a finite field, we cannot guarantee that $\omega$ actually has poles at
all points of $\cal P'$. However, the set of poles 
${\cal P}=\{P_1,\ldots ,P_n\}\subseteq\cal P'$ 
consists of $n\ge n'-g$ points. Put ${\bf P}=P_1+\ldots +P_n$. 
Of course, $\omega \in\Omega \left({\bf P}-2\cD-\cE\right)$. 
Let $w=\left( {\Res}_{P_1}\left( \omega \right) ,\ldots ,{\Res}
_{P_n}\left( \omega \right)\right)$.

The algebraic geometry code $C_L\left( X,\cD,{\cal P}\right)$
is defined as the image of the evaluation map
\[
\begin{array}{lll}
L\left( \cD\right)  & \longrightarrow  & {\Bbb F}_q^n, \\
f & \longmapsto  & \left( f\left( P_1\right) ,\ldots ,f\left( P_n\right)
\right).
\end{array}
\]

Put $C=C_L\left( X,\cD,{\cal P}\right)^\perp_w$.
For any two functions $f,g\in L\left( \cD\right) $ we have $fg\omega \in
\Omega \left( {\bf P}\right)$. Therefore $fg\omega $ has no poles except
in ${\bf P}$ and, by the residue formula, $\sum w_if\left( P_i\right)
g\left( P_i\right) =\sum {\Res}_{P_i}\left( \omega \right) =0.$ We
have proved that $C \supseteq C_w^{\perp}$.

If $q=2^m$, any element of ${\Bbb{F}}_q$ is a square, in particular, $
w_i=v_i^2$. Let $g_v$ be coordinatewise multiplication by $v=\left(
v_1,\ldots ,v_{n}\right)$. Then the code $C'=g_v\left(
C \right) $
has the property $C'\supseteq C^{\prime\perp}$.

Recall that if $a\ge 2g-1$, the parameters of $C$ and $C'$ are
\[
\begin{array}{lll}
k & = & n-a+g-1, \\
d & \geq & a-2g+2.
\end{array}
\]

Summing up, we have proved

\begin{theorem}
If there exists a curve over ${\Bbb F}_q$ of genus $g$ 
with at least $n'\ge 4g\;$  
${\Bbb F}_q$-points, then for any 
$n\le n'-g$ and any $a=2g-1,\ldots,\frac n2+g-1$ there is an 
$\left[n,k,d\right]_q$-code $C$ with
\begin{eqnarray}
k  =  n-a+g-1, \\
d  \geq  a-2g+2,
\end{eqnarray}
such that $C \supseteq C_w^\perp$ for some $w\in 
\left( {\Bbb F}_q^{*}\right) ^n$.

Moreover, if $q$ is a power of $2$, there is such a
code with $C \supseteq C_w^\perp$.
\end{theorem}

Applying, as usual, this theorem to asymptotically good families 
of curves over ${\Bbb F}_q,$ $q$ being a square, such that
\[
\frac{\left| X\left({\Bbb F}_q\right) \right| }{g\left( X\right) }
\rightarrow \sqrt{q}-1,
\]
we get

\begin{corollary}
Let $q$ be an even power of a prime. 
Then for any 
\begin{eqnarray}
\alpha \in \left( \frac 2{\sqrt{q}-2}\;\;,\;\;\frac 12+
\frac 1{\sqrt{q}-2}\right) 
\end{eqnarray} 
there exist families of codes with asymptotic
parameters
\begin{eqnarray}
R  =  1-\alpha +\frac 1{\sqrt{q}-2}\;, \\
\delta  \geq   \alpha-\frac 2{\sqrt{q}-2}\; ,
\end{eqnarray}
with the auxiliary property $C \supseteq C_w^\perp$ for some $w\in \left(
{\Bbb F}_q^{*}\right) ^n.$

If $q$ is an even power of $2$, there exist such codes with a stronger
property $C \supseteq C^\perp$.
\end{corollary}

To construct quantum codes we need a somewhat stronger statement. Recall that 
we need a triple $C'\supset C \supset C^\perp$.

If we take two divisors $\cD'\le \cD$ then 
$C_L\left(X,\cD',{\cal P}\right)\subseteq C_L\left(X,\cD,{\cal P}\right)$
and we have the opposite inclusion for duals.
The differential form $\omega$ with the above properties, good for $\cD$ is
also good for $\cD'$. 
Taking $\cD=aP_0$ and $\cD'=a'P_0$ with $a'<a$ we prove the following 

\begin{corollary}
\label{AG}
Let $q=2^{2m}$. 
Then for any pair of real numbers $(\alpha',\alpha)$ such that
$\frac 2{2^m-2}\le\alpha'\le\alpha\le \frac 12+
\frac 1{2^m-2}$ 
there exist families of triples of $2^{2m}$-ary codes
$C'\supset C \supseteq C^\perp$
with asymptotic
parameters
\begin{eqnarray}
R'  = 1-\alpha' +\frac 1{2^m-2},\\
\delta'  \geq   \alpha'-\frac 2{2^m-2},\\
R   =  1-\alpha +\frac 1{2^m-2}, \\
\delta  \geq   \alpha-\frac 2{2^m-2}.
\end{eqnarray}
Here $R'$ signifies the asymptotic rate of codes $C'$, and $R$ and $\delta$
are asymptotic parameters of codes $C$. 
\end{corollary}

\noindent {\bf Remark} Choosing an ${\Bbb F}_q$-point $P_\infty$ and taking
$\Supp \cE=\Supp \cD=\Supp \cD'=P_0$ and 
${\cal P}'=X\left({\Bbb F}_q\right)\setminus
P_0$ we see that the above codes are polynomially
constructible. This uses, of course, a difficult theorem of Vl\u{a}du\c{t},
see \cite{kats84}, \cite{tsvl91}.

\section{Summing up: quantum codes}
\label{sec:Q}

We say that a quantum code can be constructed in polynomial time if
there exists a polynomial time algorithm  constructing  explicitly
an encoder of the code and this encoder has polynomially
many elementary quantum gates.

In \cite{cleve96} it is in fact shown that knowledge of the generator matrix
of the symplectic code $F$
(also called generating operators
of the stabilizer group of $Q_F$) suffices to construct a polynomial 
complexity encoder. Moreover this  encoder construction  is, roughly
speaking,  a sequence of Gaussian eliminations of $k\times n$  matrices
and hence it has polynomial complexity.
Any generator matrix of the code $C^\perp$ could be used to construct   
a set of generator operators of ${\cal S}$ polynomially.
Finally, it is shown in \cite{kats84}, \cite{tsvl91} 
that generator matrices of
algebraic geometry codes
described in Section 
\ref{sec:AG}
can be constructed in
polynomial time. Thus the associated quantum stabilizer codes
are also  constructible in polynomial time.

To construct an asymptotically good quantum
code $Q$ we start with a family of curves $X$ over ${\bf F}_{2^{2m}}$
with $\frac{\left| X\left({\Bbb F}_q\right) \right| }{g\left(X\right)}
\rightarrow 2^m-1$.
Each curve gives us a triple $C'\supset C \supset C^\perp$
of algebraic geometry codes $C$ over ${\bf F}_{2^{2m}}$ as described in 
Section \ref{sec:AG}.
Let $C$ be an $[n, k, d]$-code and $C'$ an $[n, k', d']$-code. 
Binary expansions of $C$ and $C'$ with respect to
a self-orthogonal basis give us a triple $D'\supset D \supset D^\perp$ of
binary codes with $n_{D'}=n_D=2mn, k_{D'}=2mk', k_D=2mk, 
d_{D'}\ge d', d_D\ge d$, cf. Section \ref{sec:NB}. 
These codes give us symplectic codes
$F$, their parameters being $[2mn,2m(k+k'),\ge \min\{d,\frac32d'\}]$.
In their turn these give us quantum stabilizer 
$[[2mn,2m(k+k'-n),\ge \min\{d,\frac32d'\}]]$-codes $Q$.
The corresponding asymptotic parameters are
\begin{eqnarray}
R_Q &= & R+R'-1 \\
\delta_Q & \ge & \min\{ \delta, \frac32\delta'\}
\end{eqnarray}
where $R$, $R'$, $\delta$ and $\delta'$ are the 
parameters of algebraic geometry ${\bf F}_{2^{2m}}$-ary codes.

It is time to use Corollary \ref{AG}. Put $\alpha'=\frac23(\alpha+\gamma)$,
where $\gamma=\frac 1{2^m-2}$ (this choice of $\alpha'$ is optimal here).
The restrictions $2\gamma\le \alpha'<\alpha\le\frac12+\gamma$ are equal to 
$2\gamma\le\alpha\le\frac12+\gamma$. 
The asymptotic parameters of the algebraic 
geometry codes are 
\begin{eqnarray}R=1-\alpha+\gamma, \\
\delta\ge\alpha-2\gamma,\\
R'=1-\frac23\alpha+\frac13\gamma,\\ 
\delta'\ge\frac23\alpha-\frac43\gamma.
\end{eqnarray}
Their binary expansions have the same $R$ and $R'$, and the estimates for 
their $\delta$ 
and $\delta'$ are divided by $2m$.
By Corollary \ref{c:st1} the parameters of the quantum codes obtained are
\begin{eqnarray}
R_Q=R+R'-1=1+\frac43\gamma-\frac53\alpha,\\
\delta_Q\ge\frac1{2m}(\alpha-2\gamma).
\end{eqnarray}
Therefore, for any $m\ge 3$ we get a polynomial bound
\begin{eqnarray}
R_Q=1-\frac2{2^m-2}-\frac{10}3m\delta_Q  \label{r(d)}
\end{eqnarray}
with the restriction
\begin{eqnarray}
\delta_Q\le \frac1{2m}\left(\frac12-\frac1{2^m-2}\right)\;,  \label{(d)}
\end{eqnarray}
i.e.,
\begin{eqnarray}
1-\frac2{2^m-2}\ge R_Q\ge\frac16-\frac13\frac1{2^m-2}\;\;. \label{R}
\end{eqnarray}

Theorem \ref{main} now follows from (\ref{r(d)}) and (\ref{R}) 
by direct computation.

On Fig.1 we present the Gilbert--Varshamov type bound (\ref{vg4}) 
and the polynomial bound of Theorem 1
based on  (\ref{r(d)}) and (\ref{R}).


\begin{figure}[htb]
\begin{center}
\begin{picture}(0,0)
\epsfig{file=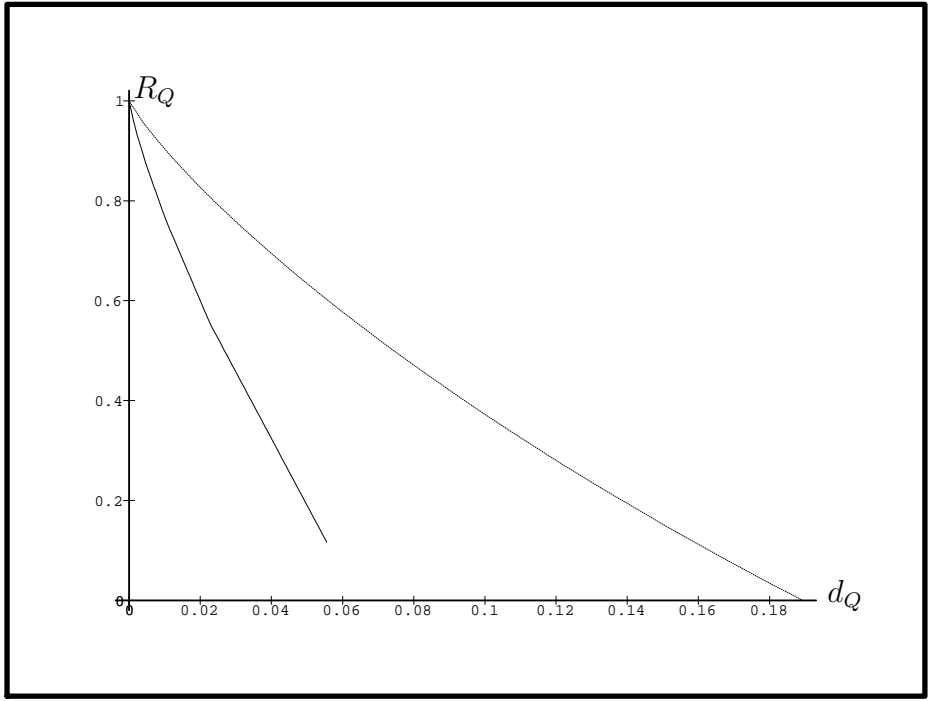}
\end{picture}
\setlength{\unitlength}{1bp}
\begin{picture}(350,280)
\put(300,30){\makebox(0,0){$d_Q$}}
\put(40,220){\makebox(0,0){$R_Q$}}
\end{picture}
\end{center}
\caption{Non-constructive bound (\ref{vg4}) and polynomial bound of
 Theorem 1}
\end{figure}

\pagebreak

\noindent
{\bf Acknowledgements}

The authors would like to thank M.Vyalyi for many fruitful discussions, and 
S.Lvovski for finding an important gap in the preliminary version of 
this paper.

\end{document}